\documentclass[twocolumn,twoside]{IEEEtran}

\usepackage{booktabs} 
\usepackage{amsmath}
\usepackage{graphicx}
\usepackage{epstopdf}
\usepackage{pseudocode}
\usepackage{enumitem}
\usepackage{color}

\title{A contemporary science map\\through the lens of IEEE and ACM periodicals}

\author{%
      George Margaritis$^{1}$, Dionysios Kritsas$^{1}$, Dimitrios Katsaros$^{1}$, and Yannis Manolopoulos$^{2}$\\
      $^{1}$ University of Thessaly, Greece; \{geormargaritis,dkritsas,dkatsar\}@uth.gr\\
      $^{2}$ University of York, Thessaloniki campus, Greece; imanolopoulos@yorkeuropecampus.eu
}

\begin{document}

\maketitle

\thispagestyle{empty}
\pagestyle{empty}

\begin{abstract}
ACM and IEEE are the two premier associations on computing and electrical/electronics engineering which publish and organize the great majority of periodicals
and conferences, respectively, serving these disciplines. Science is a constantly evolving process, and these publication fora are expected to follow the trends.
In this article, we focus on the periodicals published by the two associations and seek to detect and/or confirm any contemporary science trends as these are 
reflected to the periodical titles established recently. Our study is rather qualitative than quantitative, aiming at revealing patterns immediately 
comprehensible and validatable by the reader. Among the most notable patterns, we see a growing preference of both associations for the open access mode of 
publication; we also observe ACM's orientation toward AI-focused periodicals, and most importantly, a significant theme overlap among periodicals of the same 
association and this is valid for both ACM and IEEE.
\end{abstract}

\IEEEpeerreviewmaketitle


\section{Introduction}
\label{sec-intro}

Traditionally, scientific journals are considered as the primary means for disseminating scholarly research~\cite{Kim-Wiley-JASIST19}. Even though the situation 
is slightly different in computer science, where in some of its subfields, the conferences are considered to some extend the primary channel for publishing 
scientific results~\cite{Vardi-CACM09,Vardi-CACM10,Vrettas-JASIST15}, we can still perceive that journals comprise the long term ``scientific memory" of the 
community, and thus the preferred forum to submit research.

Journals apart from being just the fora for publishing research results, they can also be perceived as a mirror of the science evolution and its contemporary
focus. When new scientific fields are emerging, new journals are established to serve the community. New journals can also be established when a field is 
expanding too much so that the number of submitted articles can not be handled and published by current journals, or when a field is broken into very specialized 
subfields that need ``specialized" new journals. At this point, we need to mention that current publication practices have evolved so that instead of only 
journals, publishers publish magazines; we will collectively call journals and magazines as {\it periodicals}.

In this article, we aim at investigating computer and communications science and engineering current focus through the lens of periodical establishment. Even 
though there exist several publishers such as ACM, IEEE, Elsevier, Springer-Nature, Wiley etc that publish periodicals on computer and communications, in our 
study we focus exclusively on the periodicals published by ACM and IEEE, since these two are the two major computer and communication engineering and science 
associations, and typically they lead the trends, and also because their periodicals have the highest impact and respect among scholars. Our work is completely 
different from those that investigate impact and/or volume of publications per publisher, e.g., the work in~\cite{Cunningham-CACM25,Petrou-link20,Petrou-link23}, 
and in general there is no strictly related work to ours.

The major aim of the present article is to shed light on the current scientific domains in computer science and engineering as served by periodicals of the two
computing associations, namely ACM and IEEE and to record any apparent and latent patterns. In this context, the present article makes the following 
contributions:
\begin{itemize}
\item it examines which are the most popular subject categories served by the periodicals,
\item it investiagates the birth and discontinuation disctribution of the periodicals, and finally
\item it examines the degree of overlap among the topics of the periodicals.
\end{itemize}

The rest of the article is structured as follows: section~\ref{sec-def-data} describes the data that we have accumulated for study and 
section~\ref{sec-def-observations} records our observations related to the evolution of the number and
topics of the two associations' periodicals.Then, in section~\ref{sec-web-tool} we briefly introduce our online Web tool that helps navigate through the world 
of IEEE and ACM periodicals, and finally, in section~\ref{sec-conclusions} we summarize the contributions and conclude the article.

\section{Data collection and measures}
\label{sec-def-data}

In general, our data concern the title of periodicals, the year of their establishment, impact measures, and similarities among the periodicals' scope.
The aforementioned information was mainly retrieved from the digital libraries of ACM (https://dl.acm.org) and IEEE (https://ieeexplore.ieee.org), and it is
presented in the next subsections.

\subsection{Statistics of periodicals}
\label{subsec-stats-periodicals}

Firstly, we will describe some very basic information that was retrieved by these digital libraries. The first observation is that IEEE has almost five times 
more periodicals than ACM; this is expected to some extent, if we consider the broader scope of IEEE (covering computing, communications, power, etc) than the 
scope of ACM which focuses mainly on ``computing", but at a greater specialization degree (e.g., IEEE TKDE versus the set of ACM TODS and ACM TOIT and ACM TOIS).
Table~\ref{tab-num-ieee-acm-periodicals} provides the detailed numbers.

\begin{table}[!hbt]
\center
\begin{tabular}{||l|c|c||}\hline\hline
{\bf association} & {\bf journals/transactions} & {\bf magazines}\\\hline
IEEE          & 282                         & 59\\\hline
ACM           & 70                          & 0\\\hline\hline
\end{tabular}
\vspace*{.25\baselineskip}
\caption{Anatomy of our dataset.}
\label{tab-num-ieee-acm-periodicals}
\end{table}

We need to say at this point that ACM publishes practically no magazines, apart from the flagship magazine of {\it Communications of the ACM} and the newly
established {\it ACM AI Letters}. It publishes only journals/transactions. So, we choose to examine only ACM's journals/transactions, called collectively
journals, and we see that there are~$70$ such journals. On the other hand, IEEEXplore accommodates $282$~journals and $59$~magazines. Some of these periodicals, 
have stopped their operation or stopped being published by IEEE (e.g., those by IMA), some others are newsletters. So we have excluded such cases from our study, 
and we are left with the number of periodicals mentioned in Table~\ref{tab-num-ieee-acm-periodicals}.

For all IEEE's periodicals we have available their establishment year, whereas for ACM's journals, there exist three journals for which we were not able to
find their establishment year.

\subsection{Impact measures}
\label{subsec-impact-measures}

Since we aimed to somehow get a picture of how newly established periodicals are perceived by the scientific community, and also we wished to be able to contrast 
the popularity of new areas with that of the old ones via the periodicals serving them, we resorted to the use of journal impact measures, namely the
{\it Journal Impact Factor}~\cite{Garfield-Science55} or JIF for short, and the CiteScore~\cite{CiteScore26}.

Let $c_{y}$ stand for the citations received within the year ``y" by the citable items published during the years ``y-1" and ``y-2"; let also $c_{y}^{i}$ stand 
for the citations received from within year ``y-i" until the year ``y" by the citable items published during the years ``y-i" until ``y", and 
finally let $n_x$ stand for the number of citable items published by the periodical within year ``x". Then, Equations~\ref{eq-jif} and~\ref{eq-citescore} define 
the JIF and CiteScore for~2024, respectively.
\begin{align}
JIF_{2024}       &= \frac{c_{2024}}{n_{2023} + n_{2022}}\label{eq-jif}\\
CiteScore_{2024} &= \frac{c_{2024}^{3}}{n_{2024}+n_{2023}+n_{2022}+n_{2021}}\label{eq-citescore}
\end{align}

The astute reader will deduce from the aforementioned equations and the value of CiteScore can be approximated quite satisfactory by twise the value of JIF
(this can also be observed by contrasting Figure~\ref{fig-top10-IF-IEEE} to~\ref{fig-top10-CiteScore-IEEE} and Figure~\ref{fig-top10-IF-ACM} 
to~\ref{fig-top10-CiteScore-ACM}). So, in this article we mainly focus on the JIF.

Apparently, periodicals published {\it strictly after} 2021 have not received a JIF yet, and similarly, periodicals established {\it strictly after} 2020 
have not received a CiteScore yet. In Table~\ref{table-having-if-citescore} we show for how many IEEE and ACM periodicals we got their JIF and/or CiteScore.

\begin{table}[!hbt]
\center
\begin{tabular}{||l|c|c|c||}\hline\hline
{\bf periodical} & {\bf cardinality} & {\bf w/o IF} & {\bf w/o CiteScore}\\\hline
IEEE             & 282+59=341        & 56           & 51\\\hline         
ACM              & 70                & 24           & 23\\\hline\hline   
\end{tabular}
\vspace*{.25\baselineskip}
\caption{Periodicals with and without JIF and CiteScore.}
\label{table-having-if-citescore}
\end{table}

ACM has~$14$ journals established after 2021 and thus they can not have a JIF, but there are~$10$ journals whose JIF is unknown. Also, there 
exist~$10$ ACM journals that do not have CiteScore because they were established in or after 2022, and there exist~$3$ journals whose
establishment year could not be found. Moreover, there exist~$7$ journals that started in or before 2021, but we lack their CiteScore.

For IEEE, there exist 56 for which their JIF is unknown, and 26 out of them were established after 2021, so they can not get a JIF yet. However, for the rest 30 
we were not able to find their JIF. On the other hand, from out 51 IEEE periodicals without CiteScore, 27 of them were established after 2021 and thus they will 
get a CiteScore in the near future.

\subsection{Periodicals' topic similarity measures}
\label{subsec-topc-similarity}

In order to assess the similarity of topics of two periodicals, since these topics are described by keywords we deployed two widely used set similarity measures,
namely the Jaccard and Dice index (or coefficient). The Jaccard coefficient with values ranging in $[0 \dots 1]$ is a statistic used for assessing the similarity 
(and thus diversity) of sets. It is defined in general by taking the ratio of two sizes, namely of the intersection size divided by the union size (see 
Equations~\ref{eq-jaccard-sim} and~\ref{eq-jaccard-dist}). Similar in spirit is the Dice coefficient defined in Equations~\ref{eq-dice-sim} and~\ref{eq-dice-dist}.

\begin{align}
J(A,B)   &= \frac{|A \cap B|}{|A \cup B|}\label{eq-jaccard-sim}\\ 
d_J(A,B) &= 1 - J(A,B)\label{eq-jaccard-dist}\\                   
D(A,B)   &= \frac{2|A \cap B|}{|A| + |B|}\label{eq-dice-sim}\\    
d_D(A,B) &= 1 - D(A,B)\label{eq-dice-dist}                        
\end{align}

To exemplify the first definition, let us take as example the following two IEEE periodicals: IEEE Transactions on Knowledge and Data Engineering (TKDE) with 
subject(s) ``Computing and Processing" and IEEE Wireless Communications magazine (WComm) with subjects ``a) Communication, Networking and Broadcast Technologies, 
b) Computing and Processing, c) Signal Processing and Analysis". Then, $Jaccard(TKDE,WComm)=\frac{1}{3}=0.33$, and thus the Jaccard distance 
equals~$d_J(TKDE,WComm)=1-0.33=0.67$.

\section{Periodicals' impact, lifetime and themes}
\label{sec-def-observations}

In the next sections, we present the main findings of our investigation concerning periodicals (and thus topics) with the largest impact, the topics that
require new periodicals and finally the overlap of periodicals' focus areas.

\subsection{The top performing periodicals}
\label{subsec-top-performers}

\begin{figure}[!htb]
  \centering
  \includegraphics[width=.99\linewidth]{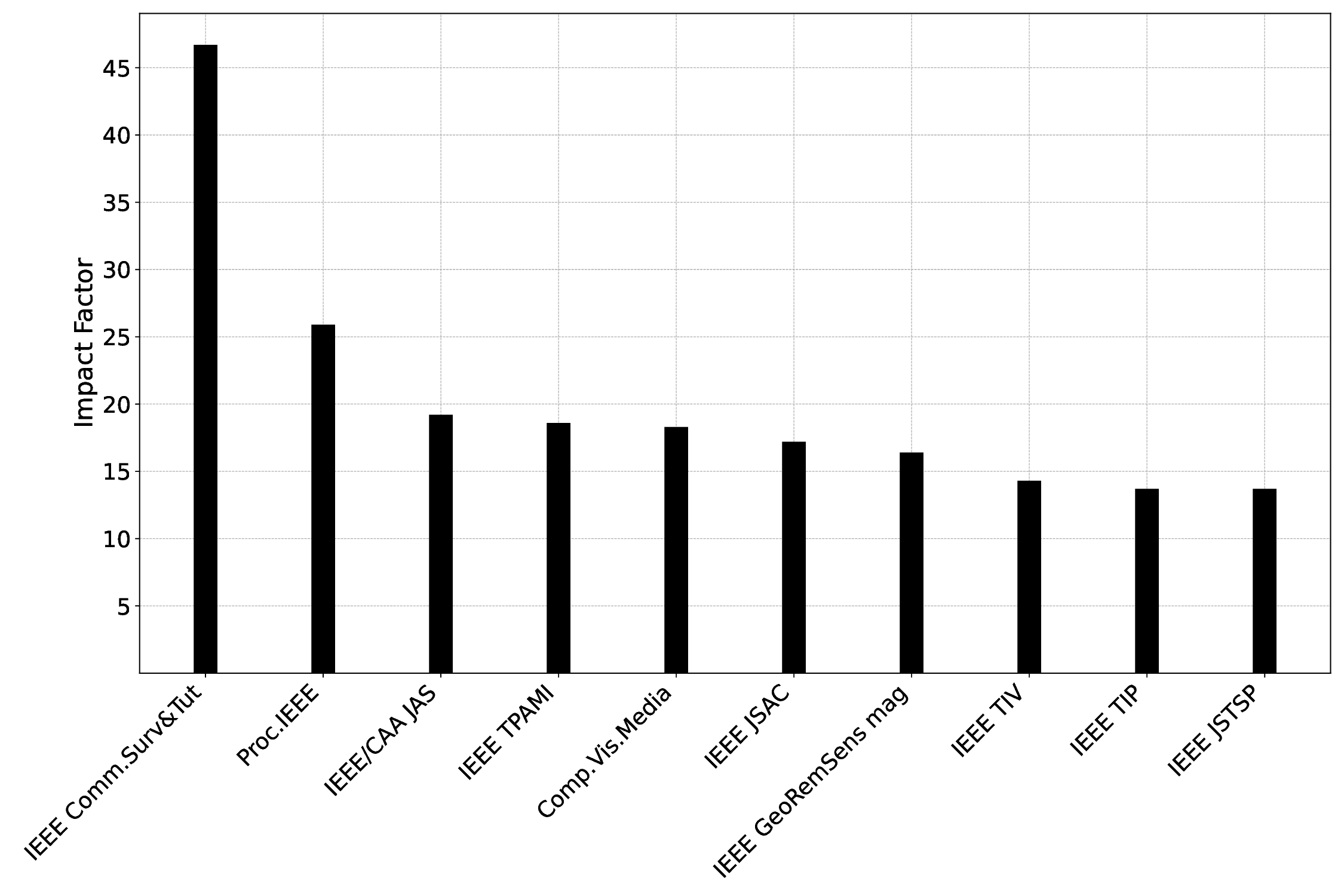}
  \caption{Top-10 IEEE journals by Impact Factor in~$2024$.}
  \label{fig-top10-IF-IEEE}
\end{figure}

\begin{figure}[!htb]
  \centering
  \includegraphics[width=.99\linewidth]{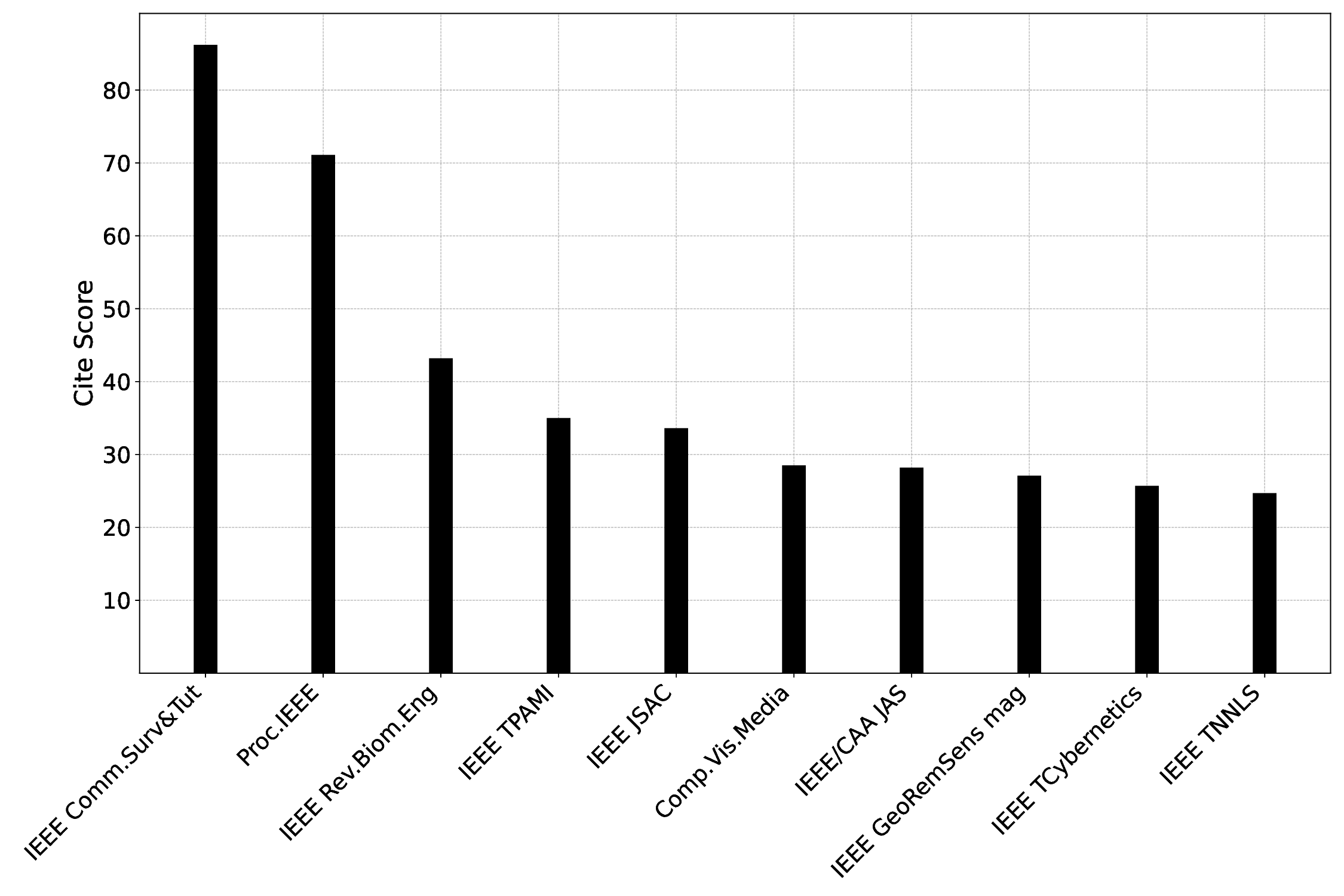}
  \caption{Top-10 IEEE journals by CiteScore in~$2024$.}
  \label{fig-top10-CiteScore-IEEE}
\end{figure}

\begin{figure}[!htb]
  \centering
  \includegraphics[width=.99\linewidth]{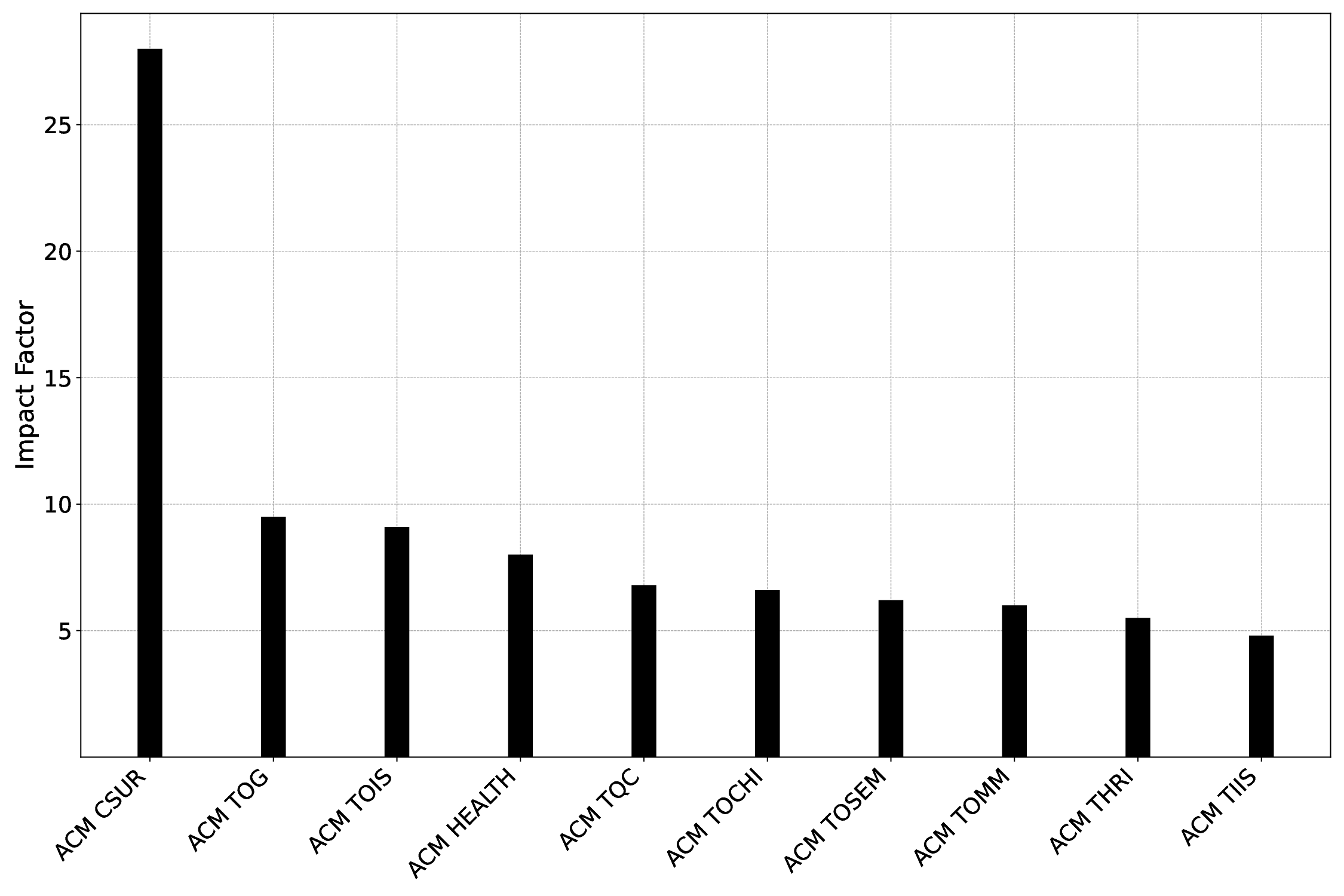}
  \caption{Top-10 ACM journals by Impact Factor in~$2024$.}
  \label{fig-top10-IF-ACM}
\end{figure}

\begin{figure}[!htb]
  \centering
  \includegraphics[width=.99\linewidth]{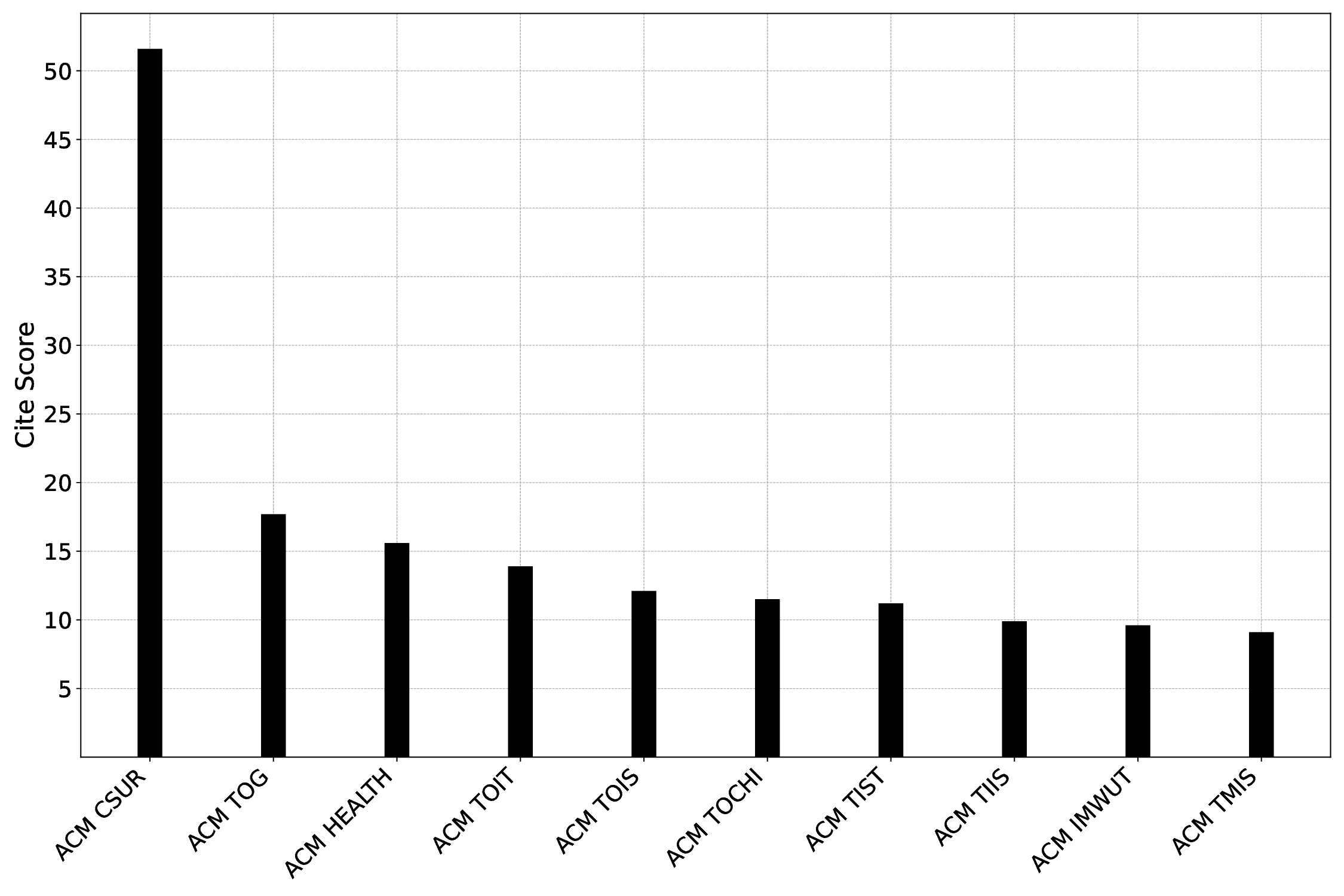}
  \caption{Top-10 ACM journals by CiteScore in~$2024$.}
  \label{fig-top10-CiteScore-ACM}
\end{figure}

The top performing journals according to Impact Factor for IEEE\footnote{In Figure~\ref{fig-top10-IF-IEEE}, IEEE JSTSP stands for IEEE Journal of Selected 
Topics in Signal Processing.} and ACM are illustrated in Figure~\ref{fig-top10-IF-IEEE} and Figure~\ref{fig-top10-IF-ACM}, respectively, whereas the top 
performing journals according to CiteSore for IEEE and ACM are illustrated in Figure~\ref{fig-top10-CiteScore-IEEE} and Figure~\ref{fig-top10-CiteScore-ACM}.

As expected, periodicals publishing survey articles, namely ACM CSUR and IEEE Communication Surveys \& Tutorials are the top performers with respect to both
JIF and CiteScore. After them, we see that the second best performing periodicals are those whose area focuses around {\it learning, intelligence, robotics} 
etc. These are the following:
\begin{itemize}
\item IEEE Transactions on Pattern Analysis and Machine Intelligence (TPAMI), 
\item IEEE/CAA Journal of Automatica Sinica (JAS), 
\item Computational Visual Media (Comp.Vis.Media)\footnote{Computational Visual Media is a fully open access journal, published bimonthly by Tsinghua University 
      Press with all articles available on IEEE Xplore from volume 11/2025.}, 
\item IEEE Transactions on Intelligent Vehicles (TIV),
\item IEEE Transactions on Image Processing (IEEE TIP), 
\item ACM Transactions on Intelligent Systems and Technology (TIST), 
\item ACM Transactions on Interactive Intelligent Systems (TIIS), etc.
\end{itemize}

On the other hand, ACM's most impactful periodicals do not focus on any specific topic, but we can see them ranging from graphics, health, software 
engineering, multimedia, intelligent systems, etc.

\begin{figure}[!htb]
  \centering
  \includegraphics[width=.99\linewidth]{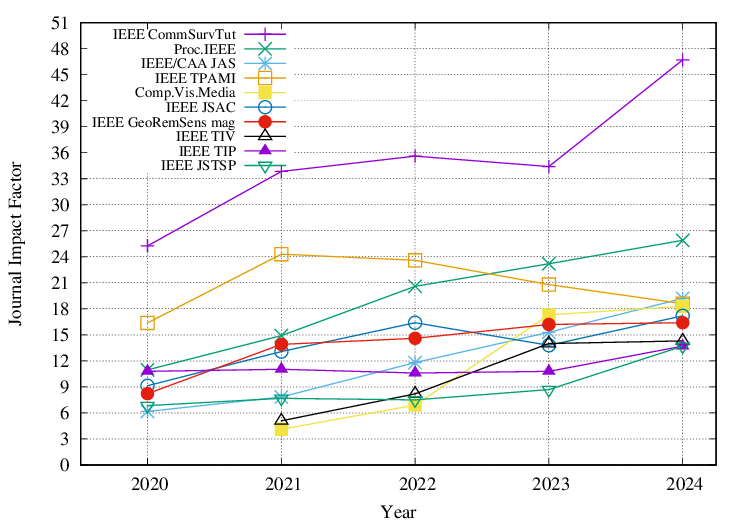}
  \caption{Impact evolution of those IEEE periodicals with the highest JIF in~$2024$.}
  \label{fig-JIF-top10-ieee-evolution}
\end{figure}

The next question that we would like to investiagte is whether these IEEE and ACM periodicas have always been on the top-$10$ and what was the evolution of their
impact. In Figure~\ref{fig-JIF-top10-ieee-evolution} we illustrate the impact evolution of those IEEE periodicals with the highest JIF for~$2024$. 

So, we see that these periodicals are building their impact/reputation linearly within the last five years, with the sole exception of IEEE TPAMI whose impact 
declines steadily since~$2021$.

\begin{figure}[!htb]
  \centering
  \includegraphics[width=.99\linewidth]{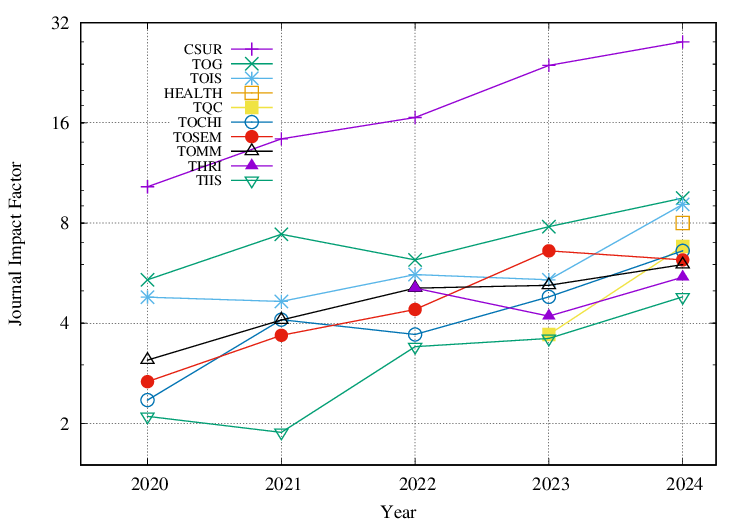}
  \caption{Impact evolution of those ACM periodicals with the highest JIF in~$2024$.}
  \label{fig-JIF-top10-acm-evolution}
\end{figure}

As far as the ACM periodicals which are toppers in~$2024$, we see that their impact increases twofold since~$2020$, with the exception of ACM CSUR which presents
a three-fold increase (see Figure~\ref{fig-JIF-top10-acm-evolution}).

Let us present now what was the IEEE periodicals with the largest JIF in a sample of recent years. 
In Tables~\ref{table-top10-jif-2023-ieee}--\ref{table-top10-jif-2020-ieee}, we present the toppers with respect to JIF for the years~$2023$, $2022$ and $2020$.
Taking into account these tables and Figure~\ref{fig-top10-IF-IEEE} and focusing on the ranking rather than absolute values of JIF, we can deduce that the 
{\it impact of periodicals serving communications and signals gradually reduces}, whereas {\it periodicals serving intelligence gain impact}.

\begin{table}[!hbt]
\center
\begin{tabular}{||l|c||}\hline\hline
IEEE Communications Surveys \& Tutorials	              & 34.4\\\hline
Proceedings of the IEEE	                                  & 23.2\\\hline
IEEE Trans.\ on Pattern Analysis and Machine Intelligence & 20.8\\\hline
IEEE Reviews in Biomedical Engineering	                  & 17.2\\\hline
IEEE-CAA Journal of Automatica Sinica	                  & 15.3\\\hline
IEEE Geoscience and Remote Sensing Magazine	              & 16.2\\\hline
IEEE Transactions on Intelligent Vehicles	              & 14.0\\\hline
IEEE Journal on Selected Areas in Communications	      & 13.8\\\hline
IEEE Transactions on Industrial Informatics	              & 11.7\\\hline
IEEE Transactions on Evolutionary Computation	          & 11.7\\\hline\hline
\end{tabular}
\vspace*{.25\baselineskip}
\caption{Top-10 IEEE journals by Impact Factor in~$2023$.}
\label{table-top10-jif-2023-ieee}
\end{table}

\begin{table}[!hbt]
\center
\begin{tabular}{||l|c||}\hline\hline
IEEE Communications Surveys \& Tutorials                  & 35.6\\\hline
Proceedings of the IEEE                                   & 20.6\\\hline
IEEE Trans.\ on Pattern Analysis and Machine Intelligence & 23.6\\\hline
IEEE Reviews in Biomedical Engineering                    & 17.6\\\hline
IEEE Journal on Selected Areas in Communications          & 16.4\\\hline
IEEE Signal Processing magazine                           & 14.9\\\hline
IEEE Geoscience and Remote Sensing Magazine               & 14.6\\\hline
IEEE Transactions on Evolutionary Computation             & 14.3\\\hline
IEEE Wireless Communications magazine                     & 12.9\\\hline
IEEE Transactions on Industrial Informatics               & 12.3\\\hline\hline
\end{tabular}
\vspace*{.25\baselineskip}
\caption{Top-10 IEEE journals by Impact Factor in~$2022$.}
\label{table-top10-jif-2022-ieee}
\end{table}

\begin{table}[!hbt]
\center
\begin{tabular}{||l|c||}\hline\hline
IEEE Communications Surveys \& Tutorials                  & 25.24\\\hline
IEEE Trans.\ on Pattern Analysis and Machine Intelligence & 16.38\\\hline
IEEE Trans. on Systems, Man and Cybernetics: Systems      & 13.45\\\hline
IEEE Signal Processing magazine                           & 12.55\\\hline
IEEE Transactions on Fuzzy Systems                        & 12.02\\\hline
IEEE Wireless Communications magazine                     & 11.97\\\hline
IEEE Transactions on Cybernetics                          & 11.44\\\hline
IEEE Computational Intelligence magazine                  & 11.35\\\hline
Proceedings of the IEEE                                   & 10.96\\\hline
IEEE Transactions on Image Processing                     & 10.85\\\hline\hline
\end{tabular}
\vspace*{.25\baselineskip}
\caption{Top-10 IEEE journals by Impact Factor in~$2020$.}
\label{table-top10-jif-2020-ieee}
\end{table}

\subsection{The birth and discontinuation of periodicals}
\label{subsec-birth-death}

\begin{figure}[!htb]
  \centering
  \includegraphics[width=.99\linewidth]{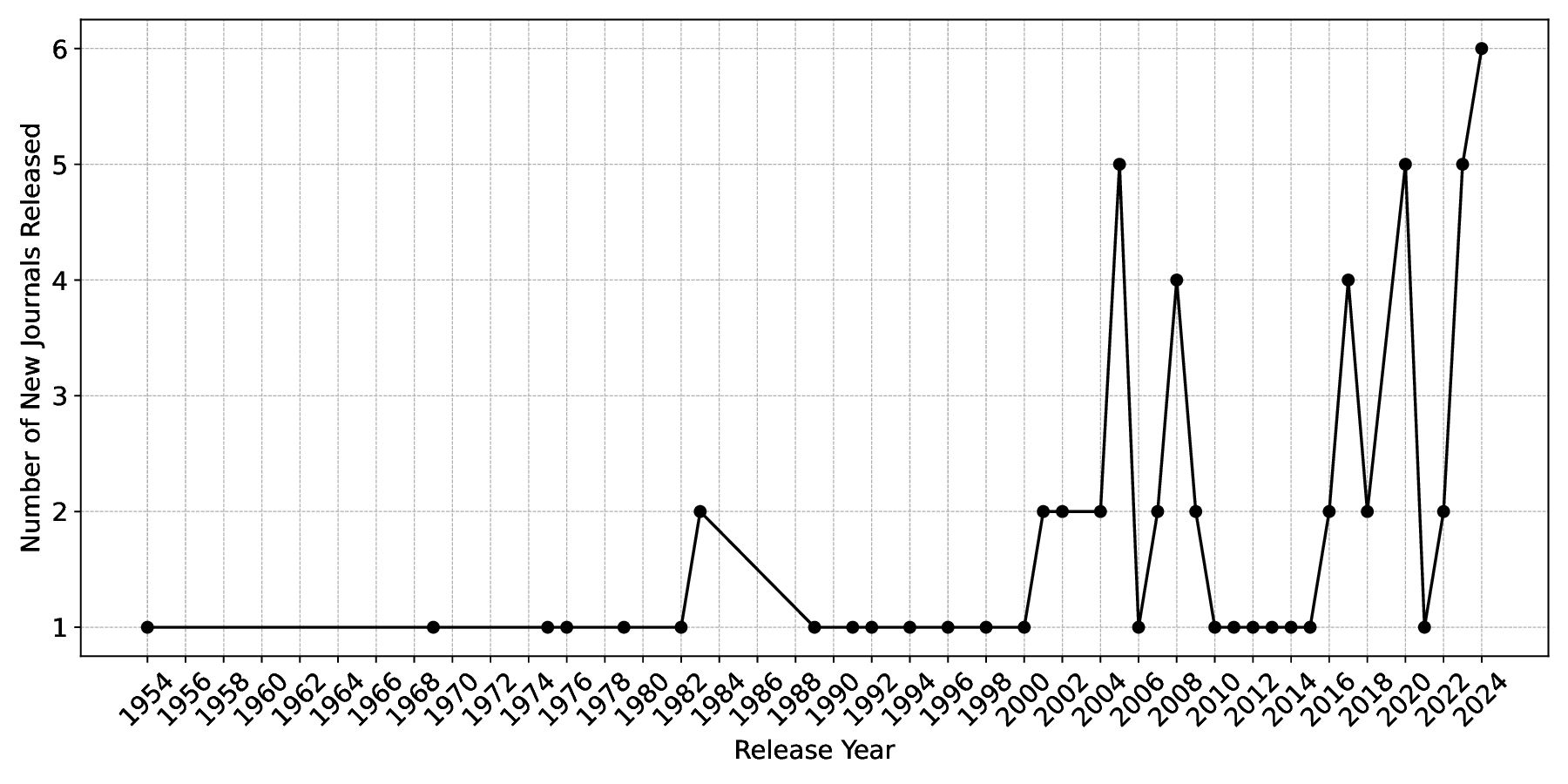}
  \caption{Birth distribution of ACM journals.}
  \label{fig-birth-jnls-ACM}
\end{figure}

The birth rates for ACM journals are depicted in Figure~\ref{fig-birth-jnls-ACM}. After 2020, we can note a couple of peaks in years~$2023$ and~$2024$, where 
ACM established the following journals: 
\begin{itemize}
\item 2023
  \begin{itemize}
  \item ACM Journal on Computing and Sustainable Societies, 
  \item ACM Transactions on Recommender Systems,
  \item Games: Research and Practice, 
  \item Proceedings of the ACM on Management of Data, 
  \item Proceedings of the ACM on Networking, 
  \end{itemize}
\item 2024
  \begin{itemize}
  \item ACM Journal on Autonomous Transportation Systems, 
  \item ACM Journal of Data Science, 
  \item ACM Journal on Responsible Computing, 
  \item ACM/IMS Journal of Data Science, 
  \item ACM  Transactions on Probabilistic Machine Learning.
  \item Proceedings of the ACM on Software Engineering, and 
  \end{itemize}
\end{itemize}

So, we observe that the majority of these new periodicals focus on data management and ``learning", i.e., data science, machine learning, recommendation, 
autonomous systems. This trends goes on for subsequent years, when ACM establishes some more new periodicals, namely ACM Transactions on AI for Science, ACM 
Transactions on AI Security and Privacy, and ACM AI Letters. Therefore, it seems that {\it ACM has a clear orientation towards establishing new journals that 
focus on the generic area of artificially intelligence and intelligent systems}.

\begin{figure}
  \centering
  \includegraphics[width=.99\linewidth]{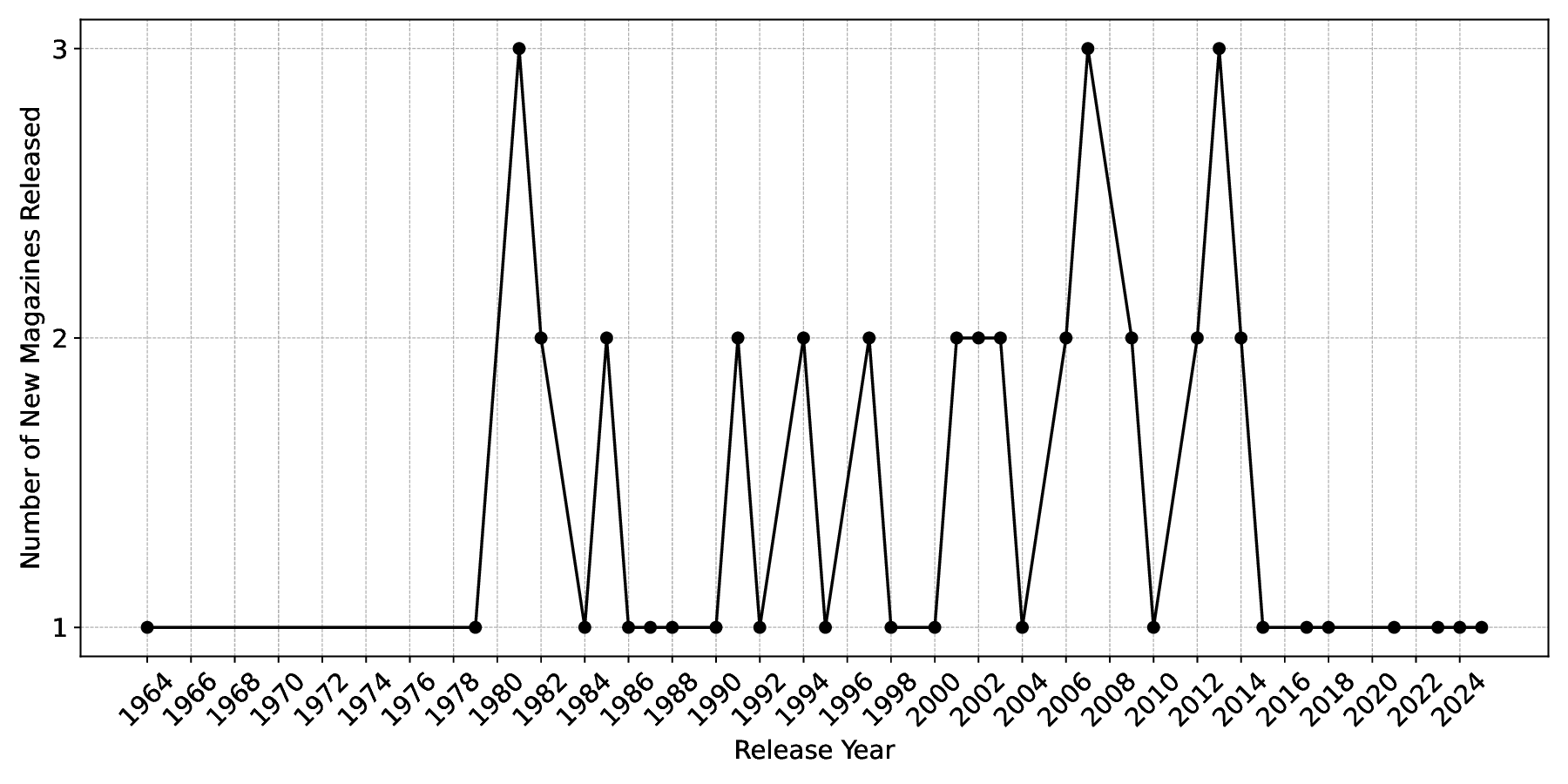}
  \caption{Birth distribution of IEEE magazines.}
  \label{fig-birth-mags-IEEE}
\end{figure}

\begin{figure}
  \centering
  \includegraphics[width=.99\linewidth]{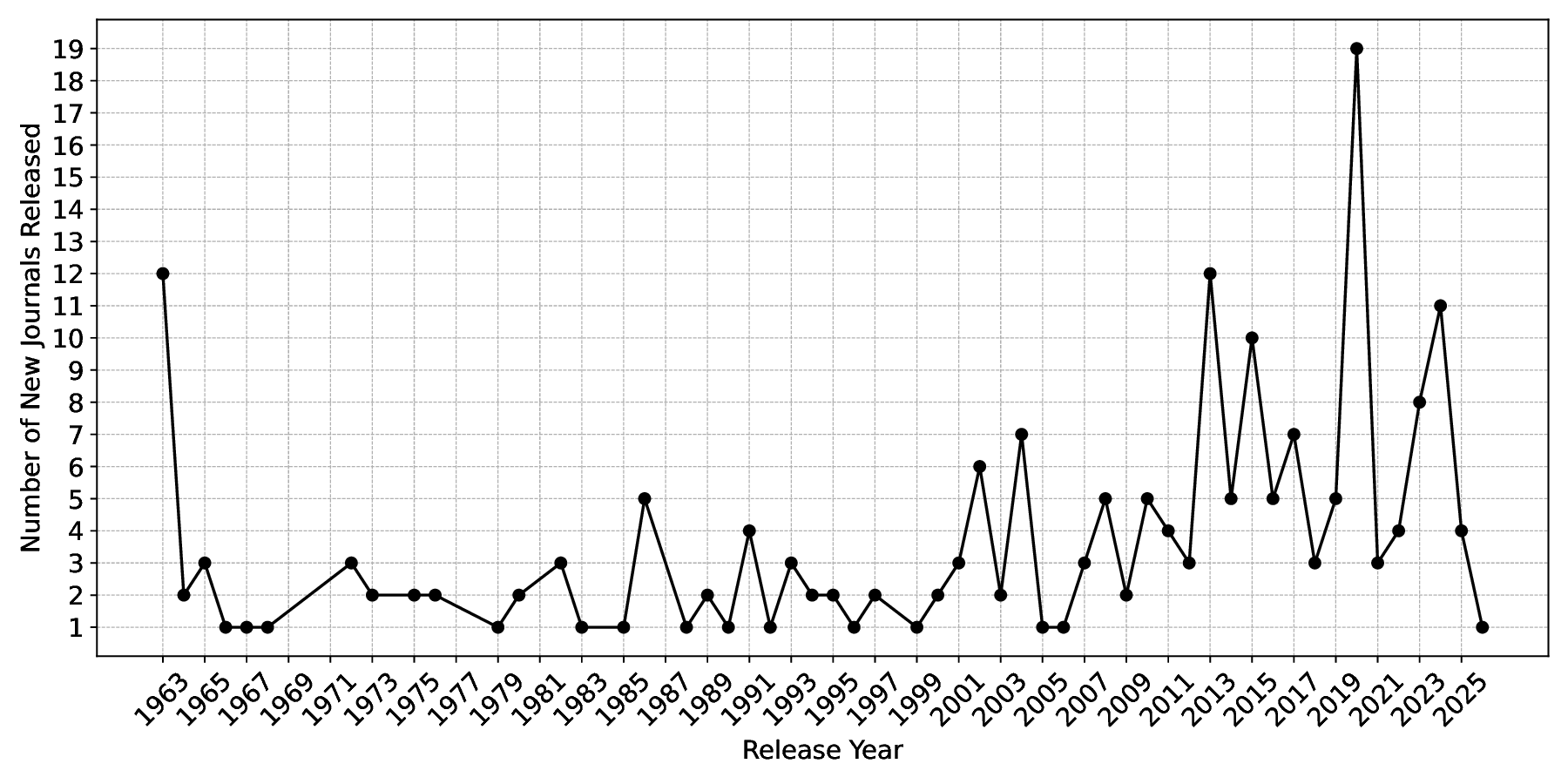}
  \caption{Birth distribution of IEEE journals.}
  \label{fig-birth-jnls-IEEE}
\end{figure}

The birth rates for IEEE magazines and journals are illustrated in Figure~\ref{fig-birth-mags-IEEE} and Figure~\ref{fig-birth-jnls-IEEE}, respectively. 
As far as the magazines are concerned, we note that establishment of new ones is not very frequent. During the past five years, only four new magazines
are published, namely 
\begin{itemize}
\item IEEE BITS Information Theory Magazine (2021), 
\item IEEE Electron Devices Magazine (2023), 
\item IEEE Reliability Magazine (2024), and
\item IEEE Energy Sustainability Magazine (2025), 
\end{itemize}
without any particular focus on some field.

But, when observing the establishment of new IEEE journals, we see a couple of peaks in years~$2020$ and~$2024$, when~$19$ and~$11$ new journals appear 
in these years, respectively. In particular, $14$~out of~$19$ new journals in year 2020, are ``Open Journal of" designating a strong the transition to the
open access model\footnote{A model that is completely followed by ACM, starting from 2026.}. The Appendix contains a detailed presentation of the new titles for 
these years. Apart from this, we can not see any other strong pattern in the newly established IEEE titles, meaning that we see {\it new IEEE periodicals related
to electronics, to signals, to systems, to AI, and so on, in a quite balanced way}.

\begin{figure}[!htb]
  \centering
  \includegraphics[width=.99\linewidth]{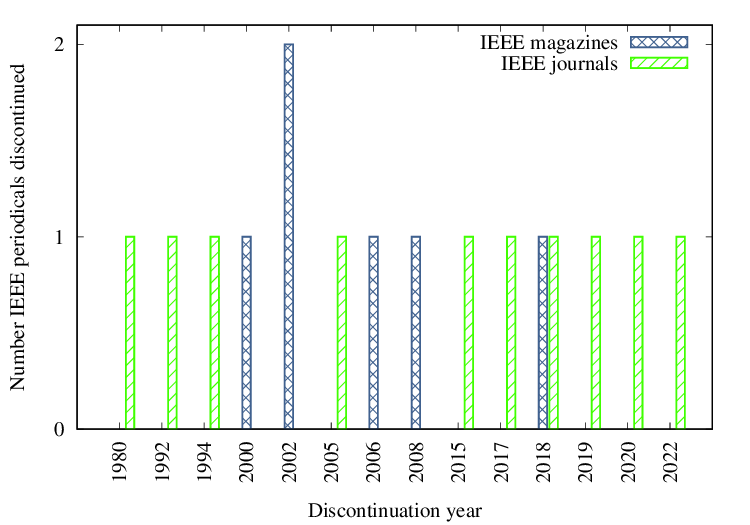}
  \caption{Discontinuation distribution of IEEE periodicals.}
  \label{fig-death-alls-IEEE}
\end{figure}

Then, we looked at the discontinuation distribution, namely we asked whether periodicals `die' as a consequence of reduced interest in an area. We present the
results in Figure~\ref{fig-death-alls-IEEE} for both journals and magazines.

Looking at the recent past, only one magazine ceased its operation, namely the IEEE Cloud Computing magazine in~$2018$. This seems somewhat irrational, since 
cloud computing continues to be a very hot research and development area especially because of the progress in AI, and moreover because this area is supported 
by a very vibrant community of researchers.

On the other hand, examining the discontinuation of IEEE journals in the recent past, we will see that only five journals stopped being published (the year of 
discontinuation in parenthesis), namely
\begin{itemize}
\item IEEE Transactions on Multi-Scale Computing Systems (2018),
\item IEEE RFID Virtual Journal (2018), 
\item IEEE Biometrics Compendium (2019), 
\item IEEE Letters of the Computer Society (2020), and 
\item IEEE RFIC Virtual Journal (2022).
\end{itemize}

From among them, only the first is an ``actual" journal, whereas the others are newsletters or do not publish original material. 
For instance, from 2010 until 2019, the IEEE Biometrics Council has published the IEEE Biometrics Compendium which was a selection of relevant biometrics related 
articles from IEEE Transactions on Image Processing, IEEE Transactions on Information Forensics and Security and IEEE Transactions on Systems, Man, and 
Cybernetics, Part B (Cybernetics). Therefore, {\it only one IEEE journal stopped recently its operation}.

\subsection{Topic categories}
\label{sebsec-topic-categories}

To examine the most popular scientific disciplines covered by ACM's and IEEE's periodicals, we present in Table~\ref{tab-acm-topics} and~\ref{tab-ieee-topics}
the top-$10$ categories served by the periodicals. The number next to each catergory records the percentage of periodicals that serve this specific category. 
Apparently, a category (subject) might appear in several periodicals, so the percentages add up to more than~$100$\%. This is also an indication of overlap among
the topics of periodicals (cf.\ section~\ref{subsec-periodical-overlap}).

So, {\it the most popular discipline covered by ACM's periodicals falls within the broader area of artificial intelligence (machine/deep learning)}; the second 
most signigifcant area is that of human-computer interaction. It is interesting to note that around $22$\% of periodicals accept articles relevant to natural 
language processing, and we can speculate that the huge growth of a deep learning technology, namely Foundation Models comprises the main reason.

\begin{table}[!htb]
\centering
\begin{tabular}{||l|c||}\hline\hline
{\bf Topic}                       & {\bf percent}\\\hline\hline
Machine learning                  & 57.75\\\hline
Neural networks                   & 53.52\\\hline
Artificial intelligence           & 45.07\\\hline
Embedded systems                  & 30.99\\\hline
Surveys and overviews             & 23.94\\\hline
Human computer interaction        & 22.54\\\hline
Natural language processing       & 22.54\\\hline
Design and analysis of algorithms & 21.13\\\hline
Empirical studies in HCI          & 19.72\\\hline
User studies                      & 19.72\\\hline\hline
\end{tabular}
\vspace*{.25\baselineskip}
\caption{ACM top-10 categories and the percentage of periodicals servicing this category.}
\label{tab-acm-topics}
\end{table}

IEEE's periodicals most popular category is `Computing and Processing', which certainly includes topics such as AI and machine/deep learning\footnote{IEEE does 
not have a specific subject category for AI, and periodicals serving this area declare as their subject the `Computing and Processing' area.} but it is 
significantly broader covering also data science/management; even periodicals focusing on communications include this subject in their topics. As expected, the 
next most popular category is the networking and communications technologies, followed by circuits and systems and then signals.

\begin{table}[!htb]
\centering
\begin{tabular}{||l|c||}\hline\hline
{\bf Topic}                                          & {\bf percent}\\\hline\hline
Computing and Processing                             & 50.44\\\hline
Communication, Networking and Broadcast Technologies & 33.14\\\hline
Components, Circuits, Devices and Systems            & 32.26\\\hline
Signal Processing and Analysis                       & 22.29\\\hline
General Topics for Engineers                         & 19.94\\\hline
Power, Energy and Industry Applications              & 18.18\\\hline
Fields, Waves and Electromagnetics                   & 14.66\\\hline
Robotics and Control Systems                         & 12.02\\\hline
Bioengineering                                       & 9.97\\\hline
Engineered Materials, Dielectrics and Plasmas        & 9.97\\\hline\hline
\end{tabular}
\vspace*{.25\baselineskip}
\caption{IEEE top-10 categories and the percentage of periodicals servicing this category.}
\label{tab-ieee-topics}
\end{table}

As noted earlier, summing the top percentages add up to more than $200$\% which implies that there significant overlap among the areas served by the periodicals;
so in the next section we investigate this overlap.

\subsection{Theme overlap among periodicals}
\label{subsec-periodical-overlap}

Our final effort aimed at investigating the degree of overlap among the themes of periodical of the two associations. Towards this goal we used the overlap
measures defined in Subsection~\ref{subsec-topc-similarity}. These measures are quite similar, but not identical and for the sake of completeness we include 
both of them in our presentation. The relevant figures should be interpreted in the following way: a mass concentrated towards the far right end of the plot
(i.e., close to~$1$) implies a very low overlap among the topics of the periodicals; when a significant part of the mass is concentrated around the middle
of the plot (at point~$0.5$ in $x$-axis) implies that there is a significant overlap among the topics of the periodicals.

\begin{figure}
  \centering
  \includegraphics[width=.99\linewidth]{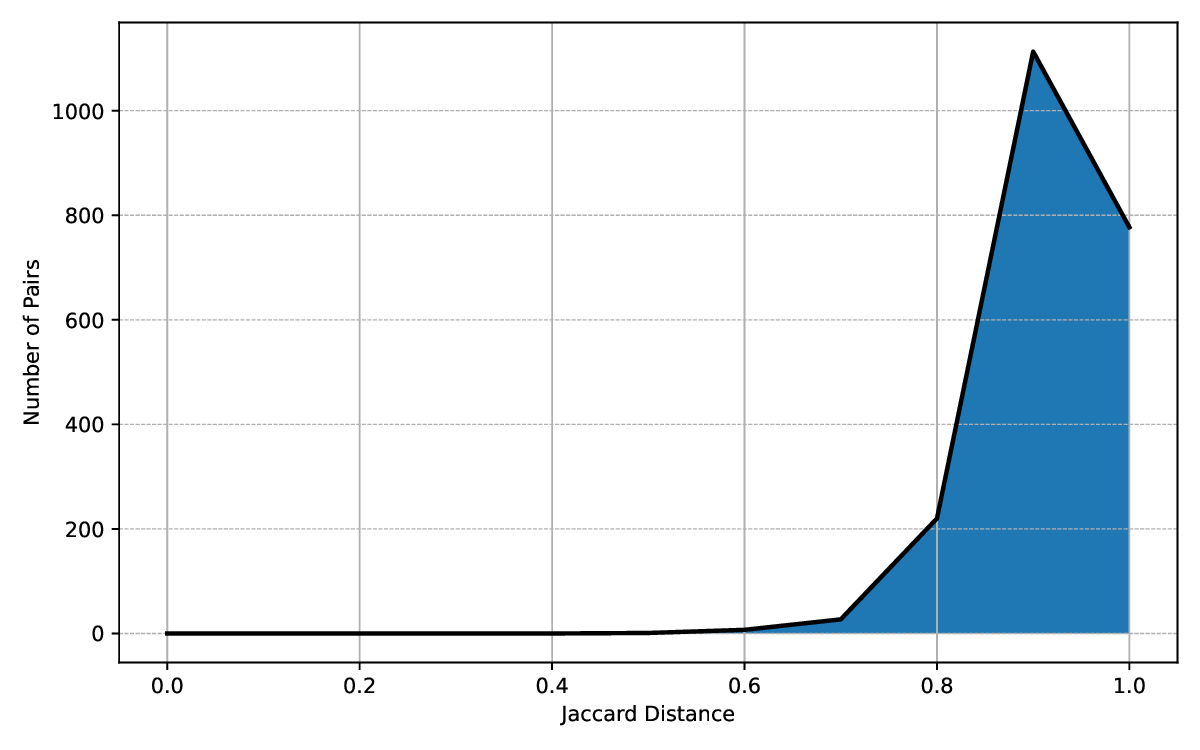}
  \caption{Distance distribution based on Jaccard coefficient for ACM journals.}
  \label{fig-dist-distr-Jaccard-ACM}
\end{figure}

\begin{figure}
  \centering
  \includegraphics[width=.99\linewidth]{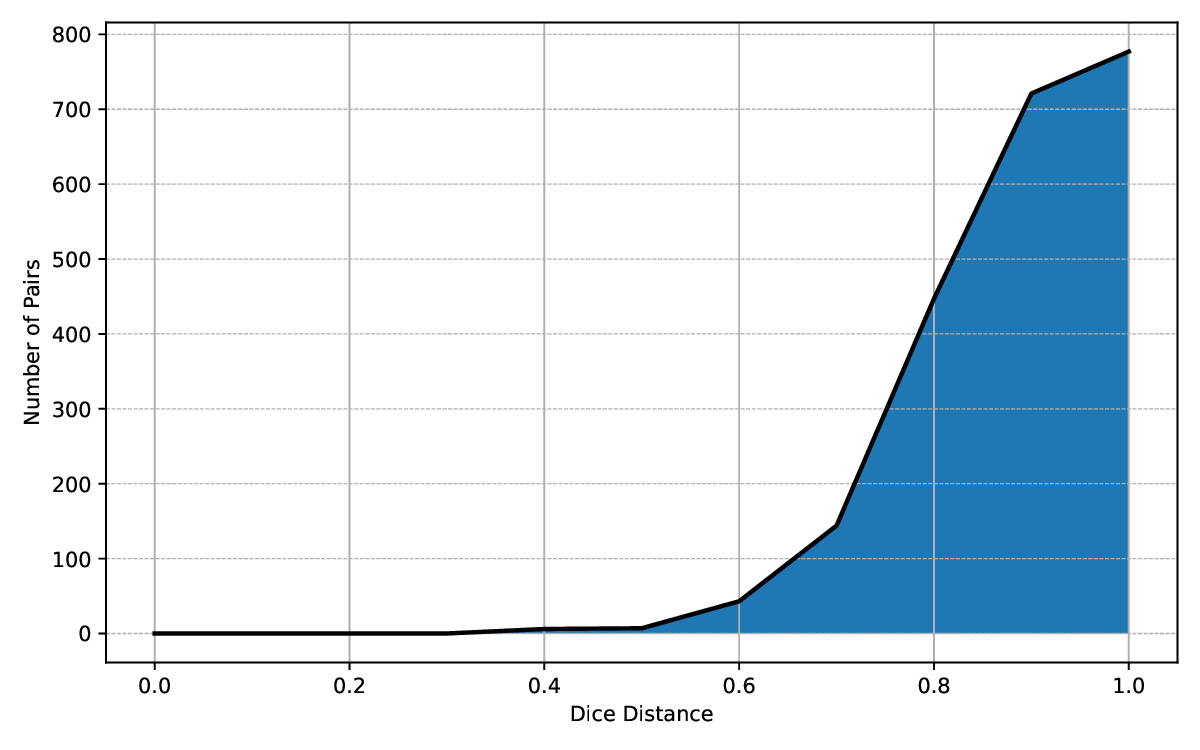}
  \caption{Distance distribution based on Dice coefficient for ACM journals.}
  \label{fig-dist-distr-Dice-ACM}
\end{figure}

Examining firstly the ACM's periodicals (the results appear in Figures~\ref{fig-dist-distr-Jaccard-ACM} and ~\ref{fig-dist-distr-Dice-ACM}) we observe that the
``center of mass" is located close to~$0.9$ (for both the Jaccard and the Dice measure). So, it seems that the ACM periodicals have a very low theme
overlap among them. There is one straightforward explanation for that and one that is more subtle. As far as the ACM periodicals are concerned, their area of 
focus is determined with a quite extended set of keywords. So, the cardinality of the keyword sets used by Jaccard (and Dice) is quite large and consequently
these large sets are not practically feasible to have significant overlap. The subtle explanation seems to be based on the ACM's tradition to have only one 
journal for a quite broad discipline, e.g., one journal for databases, one for networking, one for information systems, and so on.

\begin{figure}[!htb]
  \centering
  \includegraphics[width=.99\linewidth]{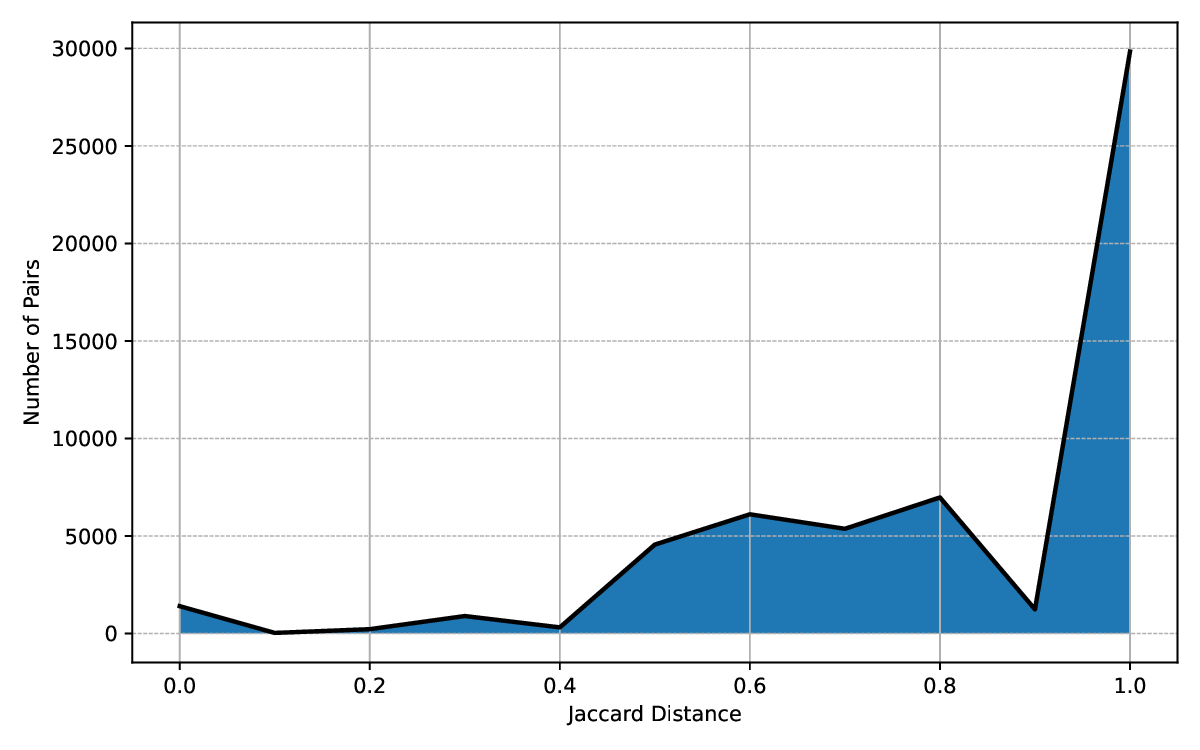}
  \caption{Distance distribution based on Jaccard coefficient for IEEE periodicals.}
  \label{fig-dist-distr-Jaccard-IEEE}
\end{figure}

\begin{figure}[!htb]
  \centering
  \includegraphics[width=.99\linewidth]{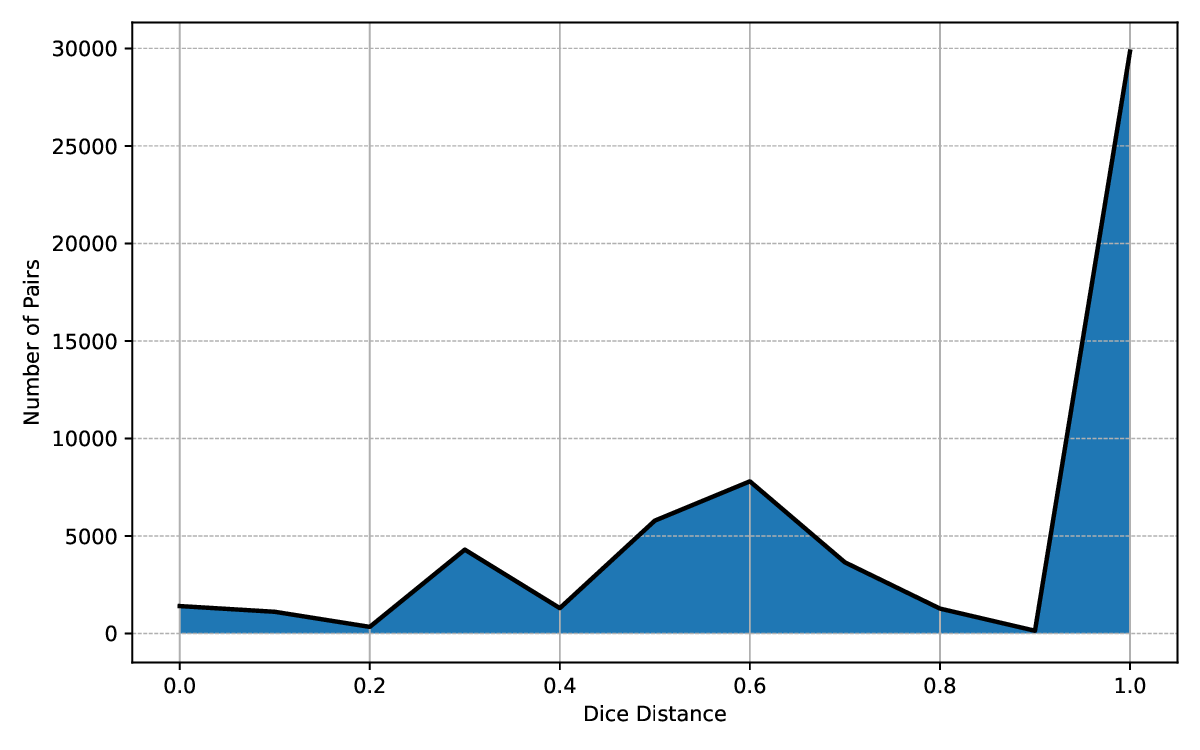}
  \caption{Distance distribution based on Dice coefficient for IEEE periodicals.}
  \label{fig-dist-distr-Dice-IEEE}
\end{figure}

The overlap profile among IEEE periodicals is apparently different. The results are depicted in Figures~\ref{fig-dist-distr-Jaccard-IEEE} 
and~\ref{fig-dist-distr-Dice-IEEE}. So, we observe that the ``center of mass" is located around~$0.7$ for the Jaccard coefficient and even more to the left
for the Dice coefficient. So, we can safely conclude that the {\it IEEE periodicals exhibit is significant overlap in their themes}. 

There seem to exist three main reasons that explain this observation. The first straightforward reason has to do with the set of keywords that describe the area 
of interest of an IEEE periodical. IEEE use a very limited set of keywords. For instance, `IEEE Transactions on Knowledge and Data Engineering (TKDE)' and `IEEE 
Wireless Communications magazine (WComm)' -- two periodicals with apparently quite diverse focus -- are described by: for TKDE the subject(s) are the following 
a) Computing and Processing for TKDE, whereas for IEEE WComm the subjects are the following: a) Communication, Networking and Broadcast Technologies, 
b) Computing and Processing, c) Signal Processing and Analysis. A second straighforward reason is that some IEEE societies publish practically a ``Transactions", 
a magazine, and a ``Letters" periodical which all have identical subjects, e.g, IEEE Transactions on Wireless Communications, IEEE Wireless Communications magazine 
and IEEE Wireless Communication Letters. However, the contribution to the aggregrate overlap due to this, is far less that the previous reason, because there are 
not too many societies that follow this pattern.

There is also a subtle reason for observing the significant overlap. This has to do with the actual areas served by the periodicals, i.e., different periodicals
serving the very same parts of a discipline. As a characteristic example consider the following periodicals ``IEEE Transactions on Communications", ``IEEE
Transactions on Wireless Communications", ``IEEE Journal on Selected Areas in Communications", ``IEEE Open Journal of the Communications Society", which all 
publish articles on communications (and networking).

\section{Web tool for navigating in the periodicals}
\label{sec-web-tool}

The investigation covered in this article was supported by the development of a publicly available Web tool found in the location https://prism.e-ce.uth.gr/.
This tool, namely PRISM (Periodical Research \& Insights for Scholarly Metrics) records the impact measures of periodicals, the subject categories, the 
establishment year of each periodical, and the overlap of pairs of periodicals. It is a searchable service, including filters to detect specific periodicals 
or pairs of them.

\section{Discussion and conclusions}
\label{sec-conclusions}

Modern science evolves very rapidly with new disciplines arising and others declining in research activity. Publication fora, namely periodicals and conferences 
need to support this continuous progress, and also to serve a steadily growing number of researchers. This article examines the periodicals i.e., journals and 
magazines published by ACM and IEEE since these are the two computing associations that currently are shaping the field, and in particular our work focuses on 
the last five years. The aim is to seek for qualitative patterns that confirm or refute a close match to the contemporary computer science and engineering 
evolution.

In the context of this question, our first objective was to identify the (top-10) most impactful periodicals. As expected, the periodicals publishing survey 
articles were on the top of the impact scale for both ACM and IEEE. Their performance keeps growing during the last five years and the performance gap with 
respect to the second best performing periodical widens. The exponentially growing population of published articles and the lengthier references list in each 
article can obviously explain this observation. As far as the rest of the top-10, we observed that most of them increase their impact during the last five
years, with the exception of IEEE TPAMI. Most of the IEEE's periodicals in the top-10 are serving a ``data analysis and learning" discipline, i.e. they
lie in the generic area of artificial intelligence, and also with a couple lying in the area of networks and signals. On the other hand, looking at the ACM's 
top-10 list, we can not observe any such clear pattern, but it is interesting to note in the top-10 a couple of periodicals serving human-machine interaction.
As a final comment for the impact, we see that the top-10 list changes year by year, with the most notable pattern being the decline of IEEE periodicals serving
``communications, networking and signals" in favor of those serving the ``learning" area.

It is expected that a changing landscape in science would also trigger the establishment and perhaps discontinuation of periodicals. Indeed, both IEEE and 
ACM have established a significant number of new periodicals during the past five years. ACM has focused almost exclusively on the broad area of AI-related 
periodicals. On the other hand, IEEE's new periodicals cover a quite diverse field of topics ranging from communications, signals, power, devices/displays, etc.
It is interesting to note that both IEEE and ACM established periodicals to service ``vertical areas", such a health, agriculture, showing a trend for 
overspecialization in publication. Another striking characteristic is the preference for the ``open access" mode of publication. IEEE has initiated nineteen
``Open Journals". ACM, starting from 2026, has turned into a complete open access mode, following a trend that other publishers such as MDPI started
long before. Additionally, it has been established that periodicals ``never die".

Finally, as far as the theme overlap among periodicals of the same association is concerned, we observed a quite high overlap. In particular, the theme overlap
is more significant for IEEE's periodicals than for ACM's periodicals. This observation can not only be explained by the larger number of IEEE periodicals, but
also by the overspecialization of their periodicals, and the existence of multiple periodicals serving the same area.

Interesting extensions of our work would be to study which periodicals become ``archival", which are ``trendy" and also to generalize the ideas of this article
to other well-established publishers such as MDPI, Elsevier, Springer-Nature.

\appendix

\section[\appendixname~\thesection]{IEEE periodicals established in 2020--2026}

A very small number of ``new" journals are simply older ones but just renamed, e.g., the old ``IEEE/ACM Transactions on Networking" has now become ``IEEE 
Transactions on Networking", the old ``IEEE/ACM Transactions on Audio, Speech and Language Processing" has now become ``IEEE Transactions on Audio, Speech and 
Language Processing", the old ``IEEE/ACM Transactions on Computational Biology and Bioinformatics" has now become ``IEEE Transactions on Computational Biology 
and Bioinformatics".

{\footnotesize
\begin{itemize}[left=1pt, label=$\diamondsuit$]
\item 2020
  \begin{enumerate}
   \item IEEE Journal of Emerging and Selected Topics in Industrial Electronics
   \item IEEE Journal on Selected Areas in Information Theory
   \item IEEE Open Journal of Antennas and Propagation
   \item IEEE Open Journal of Circuits and Systems
   \item IEEE Open Journal of Engineering in Medicine and Biology
   \item IEEE Open Journal of Industry Applications
   \item IEEE Open Journal of Intelligent Transportation Systems
   \item IEEE Open Journal of Nanotechnology
   \item IEEE Open Journal of Power and Energy
   \item IEEE Open Journal of Power Electronics
   \item IEEE Open Journal of Signal Processing
   \item IEEE Open Journal of The Communications Society
   \item IEEE Open Journal of The Computer Society
   \item IEEE Open Journal of The Industrial Electronics Society
   \item IEEE Open Journal of The Solid-State Circuits Society
   \item IEEE Open Journal of Vehicular Technology
   \item IEEE Transactions on Artificial Intelligence
   \item IEEE Transactions on Quantum Engineering
   \item IEEE Transactions on Technology and Society
  \end{enumerate}
\item 2021
  \begin{enumerate}
   \item IEEE Canadian Journal of Electrical and Computer Engineering
   \item IEEE Journal of Microwaves
   \item IEEE Open Journal of Ultrasonics, Ferroelectrics, and Frequency Control
  \end{enumerate}    
\item 2022
  \begin{enumerate}
   \item IEEE Journal on Flexible Electronics
   \item IEEE Open Journal of Control Systems
   \item IEEE Open Journal of Instrumentation and Measurement
   \item IEEE Transactions on Signal and Power Integrity
  \end{enumerate}   
\item 2023
  \begin{enumerate}
   \item IEEE Journal of Indoor and Seamless Positioning and Navigation
   \item IEEE Microwave and Wireless Technology Letters
   \item IEEE Open Journal of Systems Engineering
   \item IEEE Transactions on AgriFood Electronics
   \item IEEE Transactions on Energy Markets, Policy and Regulation
   \item IEEE Transactions on Industrial Cyber-Physical Systems
   \item IEEE Transactions on Machine Learning in Communications and Networking
   \item IEEE Transactions on Radar Systems
  \end{enumerate}    
\item 2024
  \begin{enumerate}
   \item IEEE Data Descriptions
   \item IEEE Electron Devices Reviews
   \item IEEE Robotics and Automation Practice
   \item IEEE Journal of Selected Areas in Sensors
   \item IEEE Open Journal on Immersive Displays
   \item IEEE Sensors Reviews
   \item IEEE Systems, Man, and Cybernetics Letters
   \item IEEE Transactions on Circuits and Systems for Artificial Intelligence
   \item IEEE Transactions on Field Robotics
   \item IEEE Transactions on Materials for Electron Devices
   \item IEEE Transactions on Privacy
  \end{enumerate}
\item 2025
  \begin{enumerate}
   \item IEEE Journal of Selected Topics in Electromagnetics, Antennas and Propagation
   \item IEEE Transactions on Audio, Speech and Language Processing
   \item IEEE Transactions on Computational Biology and Bioinformatics
   \item IEEE Transactions on Networking
  \end{enumerate}
\item 2026
  \begin{enumerate}
   \item IEEE Journal on Wireless Power Technologies
  \end{enumerate}
\end{itemize}
}
\normalsize

\section[\appendixname~\thesection]{ACM periodicals established in 2020--2026}

\noindent
{\footnotesize
\begin{itemize}[left=1pt, label=$\diamondsuit$]
\item 2020
  \begin{enumerate}
   \item ACM Transactions on Computing for Healthcare
   \item ACM Transactions on Internet of Things
   \item ACM Transactions on Quantum Computing
   \item Digital Government: Research and Practice
   \item Digital Threats: Research and Practice
  \end{enumerate}
\item 2021
  \begin{enumerate}
   \item ACM Transactions on Evolutionary Learning and Optimization
  \end{enumerate}
\item 2022
  \begin{enumerate}
   \item Collective Intelligence 
   \item Distributed Ledger Technologies: Research and Practice
  \end{enumerate}
\item 2023
  \begin{enumerate}
   \item ACM Journal on Computing and Sustainable Societies
   \item ACM Transactions on Recommender Systems
   \item Games: Research and Practice
   \item Proceedings of the ACM on Management of Data
   \item Proceedings of the ACM on Networking
  \end{enumerate}
\item 2024
  \begin{enumerate}
   \item ACM Journal of Data Science
   \item ACM Journal on Autonomous Transportation Systems
   \item ACM Journal on Responsible Computing
   \item ACM Transactions on Probabilistic Machine Learning
   \item ACM/IMS Journal of Data Science
   \item Proceedings of the ACM on Software Engineering
  \end{enumerate}
\item 2026
  \begin{enumerate}
   \item ACM AI Letters 
   \item ACM Transactions on AI for Science
   \item ACM Transactions on AI Security and Privacy
  \end{enumerate}
\end{itemize}
}
\normalsize

\bibliographystyle{IEEEtran}
\bibliography{periodicals}

@ARTICLE{Cunningham-CACM25,
    author      = "Cunningham, P. and Smyth, B.",
    title       = "An analysis of the impact of {G}old {O}pen {A}ccess publications in computer science",
    journal     = "Communications of the ACM",
    volume      = "68",
    number      = "9",
    pages       = "62--69",
    year        = "2025",
}

@ARTICLE{Vardi-CACM09,
    author      = "Vardi, M.",
    title       = "Conferences vs.\ journals in computing research",
    journal     = "Communications of the ACM",
    volume      = "52",
    number      = "5",
    pages       = "5",
    year        = "2009",
}

@ARTICLE{Vardi-CACM10,
    author      = "Vardi, M.",
    title       = "Revisiting the publication culture in computing research",
    journal     = "Communications of the ACM",
    volume      = "53",
    number      = "3",
    pages       = "5",
    year        = "2010",
}

@ARTICLE{Vrettas-JASIST15,
    author      = "Vrettas, G. and Sanderson, M.",
    title       = "Conferences versus journals in computer science",
    journal     = "Journal of the Association for Information Science and Technology",
    volume      = "66",
    number      = "12",
    pages       = "2674--2684",
    year        = "2015",
}

@ARTICLE{Kim-Wiley-JASIST19,
    author      = "Kim, Jinseok",
    title       = "Author-based analysis of conference versus journal publication in computer science",
    journal     = "Journal of the Association for Information Science and Technology",
    volume      = "70",
    number      = "1",
    pages       = "71--82",
    year        = "2019",
}

@ARTICLE{Garfield-Science55,
    author      = "Garfield, E.",
    title       = "Citation indexes to science: {A} new dimension in documentation through association of ideas",
    journal     = "Science",
    volume      = "122",
    pages       = "108--111",
    year        = "1955",
}

@MISC{CiteScore26,
    author      = "Elsevier",
    title       = "Scopus CiteScore",
    year        = "2026",
    note        = "Available at: {\small https://www.elsevier.com/products/scopus/metrics/citescore/}",
}

@MISC{Petrou-link20,
    author      = "Petrou, C.",
    title       = "M{DPI}'s remarkable growth",
    year        = "2020",
    note        = "Available at: {\small https://scholarlykitchen.sspnet.org/2020/08/10/guest-post-mdpis-remarkable-growth/}",
}

@MISC{Petrou-link23,
    author      = "Petrou, C.",
    title       = "Reputation and publication volume at {MDPI} and {F}rontiers",
    year        = "2023",
    note        = "Available at: {\small https://scholarlykitchen.sspnet.org/ 2023/09/18/guest-post-reputation-and-publication-volume-at-mdpi-and-frontiers-the-1b-question/}",
}

\end{document}